# Unsupervised deconvolution of dynamic imaging reveals intratumor vascular heterogeneity and repopulation dynamics

Li Chen[1,3], Peter L. Choyke[2], Niya Wang[3], Robert Clarke[4], Zaver M. Bhujwalla[5], Elizabeth M. C. Hillman[6], Ge Wang[7] and Yue Wang[3,†]

[1]Pediatric Oncology Branch, National Cancer Institute, National Institutes of Health, Bethesda, MD 20892, USA; [2]Molecular Imaging Program, National Cancer Institute, National Institutes of Health, Bethesda, MD 20892, USA; [3]Department of Electrical and Computer Engineering, Virginia Polytechnic Institute and State University, Arlington, VA 22203, USA; [4]Lombardi Comprehensive Cancer Center, Georgetown University, Washington, DC 20057, USA; [5]Department of Radiology and Radiological Science, Johns Hopkins University School of Medicine, Baltimore, MD 21205, USA; [6]Department of Biomedical Engineering, Columbia University, New York, NY 10027, USA; [7]Department of Biomedical Engineering, Biomedical Imaging Center, Rensselaer Polytechnic Institute, Troy, NY 12180, USA

**With the existence of biologically distinctive malignant cells originated within the same tumor, intratumor functional heterogeneity is present in many cancers and is often manifested by the intermingled vascular compartments with distinct pharmacokinetics. However, intratumor vascular heterogeneity cannot be resolved directly by most *in vivo* dynamic imaging. We developed multi-tissue compartment modeling (MTCM), a completely unsupervised method of deconvoluting dynamic imaging series from heterogeneous tumors that can improve vascular characterization in many biological contexts. Applying MTCM to dynamic contrast-enhanced magnetic resonance imaging of breast cancers revealed characteristic intratumor vascular heterogeneity and therapeutic responses that were otherwise undetectable. MTCM is readily applicable to other dynamic imaging modalities for studying intratumor functional and phenotypic heterogeneity, together with a variety of foreseeable applications in the clinic.**

Intratumor genetic or epigenetic heterogeneity has been found in many cancers as evidenced by deep sequencing selectively applied to different parts of the same tumor [1,2]. Consequently, cancer cells display remarkable phenotypic variability, including ability to induce angiogenesis, seed metastases, and survive therapy [3-5]. Advanced solid tumors often contain vascular compartments with distinct pharmacokinetics, comprising hypoxic regions and spatially intermingled irregular vasculature that is leaky and inefficient [6-8]. The complexity of heterogeneity has clinical implications. A more heterogeneous tumor is more likely to fail therapy due to increased drug-resistant variants [3,5], and characteristics of the dominant cell type will not necessarily predict the behaviors of interest rooted in specific cells [4].

Dynamic contrast-enhanced magnetic resonance imaging (DCE-MRI) provides a noninvasive *in vivo* method to evaluate tumor vasculature architectures based on contrast accumulation and washout [7,9]. While DCE-MRI can potentially depict the intratumor heterogeneity of vascular permeability [10], the quantitative application of DCE-MRI has been hindered by its inability to accurately resolve vascular compartments with distinct pharmacokinetics due to limited imaging resolution [7,11]. This indistinction among the contributions of different compartments to the mixed tracer signals can confound compartment modeling and deep phenotyping for association studies [4,12,13]. The goal of the present work was to discern vascular heterogeneity and its changes in tumors using DCE-MRI and novel mathematical models, for personalized cancer diagnosis and treatment.

We developed a computational method (multi-tissue compartment modeling - MTCM) for deconvolving intratumor vascular heterogeneity and identifying pharmacokinetics changes in many biological contexts [5,14,15]. MTCM works by applying a convex analysis of mixtures that enables geometrically-principled delineation of distinct vascular structures from DCE-MRI data (**Fig. 1a-c**). A formal mathematical description of the method and its detailed implementation is available in Methods.

## Results

**Overview of MTCM.** Tumors to be analyzed by MTCM contain unknown numbers of distinct vascular compartments. The pixel-wise tracer concentration in a particular vascular compartment is modeled as being proportional to the local volume transfer constant of the vascular compartment (Method). Because there are often significant numbers of partial-volume pixels, MTCM instead estimates pharmacokinetic parameters (flux rate constants) via the time-courses of pure-volume pixels (pixels whose signal is highly enriched in a particular vascular compartment). Convex analysis of mixtures identifies those pure-volume pixels present at the vertices of the clustered pixel time series scatter simplex, without any knowledge of compartment distribution (Method). When the number of underlying vascular compartments is detected using the minimum description length (MDL) criterion, MTCM provides a completely unsupervised approach to characterize intratumor heterogeneity (**Methods** and **Appendix 1.Supplementary Discussion**).

Modeling the pharmacokinetics of each vascular compartment using pure-volume pixel time-courses allowed us to estimate individual compartment flux rate constants (**Fig. 1d-e**). Non-negative least-square estimation yielded pixel-wise local volume transfer constants (Methods and **Fig. 1f**). Using synthetic and mouse DCE-MRI experiments, we showed that MTCM can be used to estimate pharmacokinetic parameters in several vascular compartments simultaneously and to quantitatively reconstruct tissue-specific local volume transfer constants (**Supplementary Data 1-2, Fig. 3, Supplementary Fig. 2** and **Supplementary Tables 1-2**). Furthermore, MTCM enabled quantitation of differences in tissue-specific vascular permeability across time (for example, therapeutic responses in longitudinal studies; Methods). Thus, the change in values of flux rate constants in a given vascular compartment could be determined, despite an expected difference in that vascular compartment's relative abundance.

We also analyzed the same realistic synthetic dataset using a "traditional" way of principal component analysis - PCA. By a comparison of the tracer concentration extracted by PCA (Figure 3e) to that estimated by MTCM (Figure 3d), we can see that tracer concentration curves estimated by PCA are highly fluctuant and significant deviated from the ground truth. In fact, similar unsatisfactory results produced by PCA or classic factor analysis have been observed in the earlier studies by us and others (Cinotti, Bazin et al. 1991, Zhou, Huang et al. 1997, Hillman and Moore 2007, Hillman, Amoozegar et al. 2011). We should clarify that MTCM consists of two major analytic parts: convex analysis of mixtures (CAM) and compartment modeling (CM), where the CAM is a critical step that automatically identifies the pure tissue pixels and their time activity curves, followed by the CM that estimates the pharmacokinetics parameters without being contaminated by the partial-volume effect. In contrast, since PCA does not enforce the nonnegative constraint for tracer concentration estimation, a subsequent compartment modeling cannot be performed to estimate pharmacokinetic parameters.

**Intratumor vascular heterogeneity in breast cancer revealed by MTCM.** In keeping with our goal to use MTCM to better uncover vascular heterogeneity in human tumors, we applied MTCM to DCE-MRI sequence data obtained from a case of advanced breast cancer (**Fig. 1a**).

In this breast tumor [7], vascular heterogeneity is characterized by active angiogenesis in the peripheral "rim" and concurrent inner-core hypoxia. Upon preliminary analysis using MDL, we found that a two-tissue compartment model of the fast and slow tracer clearance rates was sufficient to account for the variable permeability at the majority of pixels (Methods). Thus, we used pure-volume pixels associated with these two vascular pools to estimate tissue-specific flux rate constants and to reconstruct local volume transfer constant maps (Methods). MTCM reveals two vascular compartments with distinct flux rate constants (**Fig. 1e**). Accordingly, we detected distinct spatial patterns of specific local volume transfer constant in the two vascular compartments (**Fig. 1f**) with a significant fraction of partial volume pixels.

Intratumor vascular heterogeneity identified by MTCM is consistent with the knowledge obtained from *ex vivo* microscopic and molecular studies [7,13]. Defective endothelial barrier function is one of the better documented abnormalities of tumor vessels, resulting in functional heterogeneity in vascular permeability to macromolecules [7,11]. As a tumor rapidly outgrows its blood supply, it requires neovessel maturation, often leaving an inner core of the tumor with regions where blood flow and oxygen concentration are significantly lower than in normal tissues [6]. MTCM reconstructed local volume transfer constant maps correlate well with the differential gene expression known to regulate angiogenesis [7,13].

**Changes in intratumor vascular heterogeneity in longitudinal studies.** We also detected changes in pharmacokinetic patterns among longitudinal DCE-MRI data from breast cancer acquired before, during, and after treatment (**Fig. 2a**), quantified as different flux rate constants over time (Methods and **Supplementary Table 3**). For example, the two vascular compartment time-activity curves revealed by MTCM in the baseline data are highly distinct (**Fig. 2b**). We detected significantly higher permeability in a fast-flow pool and slightly lower permeability in a slow-flow pool when compared with the normal state. In contrast, the interim response (**Fig. 2c**) exhibits vascular compartment time-activity curves that are distinct but much closer to each other, whereas the closing response (**Fig. 2d**) shows a significant decrease in permeability of the fast-flow pool. We also detected different local volume transfer constant maps (**Fig. 2b-d**) and changes in the fractions of partial-volume pixels (**Supplementary Table 4**).

**Comparative studies using standard compartment modeling.** We compared tissue-specific pharmacokinetics detected with MTCM to the results of a standard compartment analysis of (total) vascular pool within the region of interest. Total time-activity curves were indistinct across time (**Supplementary Fig. 2**) owing to therapeutic effects in some parts of the tumor but not in others and large fractions of partial-volume pixels. In this longitudinal study, we deconvolved total time-activity curves into two phased therapeutic effects using MTCM: a transient "normalization" of abnormal yet surviving tumor vasculature together with the significant and consistent drop in the relative volume transfer constants [6,16]. In contrast, standard analysis may not return informative results when both the flux rate constant and volume transfer constant change heterogeneously in response to therapy. These examples illustrate the ability of MTCM to discover intratumor vascular heterogeneity and to detect changes in each vascular compartment over time. Finally, we tested the applicability of MTCM to dynamic fluorescence imaging data acquired on a mouse after bolus injection of indocyanine green dye by deconvolving biodistribution dynamics of the major organs [17] (**Supplementary Fig. 3**). The dissected tissue compartments constitute anatomical structures of the mouse that agree well with a digital anatomical mouse atlas.

**Discussion**

Several previous studies have discussed the problem of intratumor vascular heterogeneity in compartment modelling [7,11,16,18], a major outstanding issue for the characterization of complex phenotypes and therapeutic responses. Some methods have addressed the estimation of multi-compartment pharmacokinetics in the presence of varying partial-volume effects, relying on known regions of pure-volume pixels and number of compartments [10,13,16,17]. The significant advantage of our strategy is its ability to detect and quantify intratumor vascular heterogeneity without any type of external information. The benefits of such a method include its wide applicability, sensitive detection of heterogeneity dynamics, and reliance on longitudinal data from one single subject (**Appendix 1. Supplementary Discussion**).

We have identified differential and heterogeneous changes in tissue-specific vascular pharmacokinetics in tumors during treatment that were undetected using standard analysis, including tumor islands of persistent enhancement that have escaped the effects of therapy [18]. These results are particularly intriguing when considered together with recent imaging studies describing foci of resistant and more aggressive clones within a tumor [5,13]. While it is not yet possible to assign causality, these *in vivo* results allowed us to propose new hypotheses regarding the complex relationships between intratumor heterogeneity, clonal repopulation, cancer stem-cell, and therapeutic efficacy [1,3,5,10,19].

In metastatic disease, recent studies have revealed the emergence of treatment-resistant subclones that were present at a minor frequency in the primary tumour [20]. Thus, modeling cancer diagnosis and treatment in the future should involve characterization of subpopulations within the primary tumour, monitoring of clonal dynamics during treatment and eradication of treatment-emergent clones [21]. To prospectively assess intratumor heterogeneity, profiling of multiregional tumour samples would be required. However, it is impractical and potentially risky to take multiple 'random' biopsies in every patient, owing to both sampling bias and the inability to resolve intermingled heterogeneity [22]. MTCM would not only make longitudinal in vivo surveillance possible but also enable imaging-informed selective biopsies.

The future challenges of applying MTCM lie in the gap between research experiments and clinical practice. Unlike high-quality data in well-designed research studies, clinical data are usually with limited spatial and/or temporal resolution, accompanied by higher noise level (**Supplementary Fig. 4**). Lower spatial resolution results in less pure-volume pixels and thus reduces the accuracy of MTCM; while limited temporal resolution prevents accurate differentiation and estimation of pharmacokinetic parameters associated with distinctive vascular compartments.

So far we have tested MTCM method on DCE-MRI data, dynamic contrast-enhanced optical imaging data, and dynamic PET imaging data, acquired from both human tissue/organ and whole-body mouse model (e.g., Supplementary Fig. 4). Theoretically, the MTCM method can produce confident estimation on any 'dynamic contrast-enhanced' imaging data with sufficient quality (e.g., spatial and temporal resolution) [23-25]. However, we should emphasize that there are a few fundamental assumptions behind the MTCM methodology, as specified in the newly proved theorems (e.g., linear convex combination, existence of pure-tissue pixels). As in most medical imaging analysis, object motion constitutes a major source of error and can significantly confound the modeling results. Currently, MTCM is limited to 'parallel' compartment models, while the CAM part of the MTCM algorithm is applicable to resolving partial-volume contamination problem independent of the compartment models being used for subsequent parameter estimation.

**Methods**

**Multi-tissue compartment modeling of DCE-MRI series.** Let us consider $J$-tissue compartment model of DCE-MRI series (the $J$th tissue compartment corresponds to tracer plasma input, indexed by $p$), whose tracer concentration kinetics are governed by a set of first-order differential equations (**Fig. 1c**) [26,27]

$$\frac{dC_1(t)}{dt} = K_1^{\text{trans}} C_p(t) - k_{\text{ep},1} C_1(t),$$
$$\vdots$$
$$\frac{dC_{J-1}(t)}{dt} = K_{J-1}^{\text{trans}} C_p(t) - k_{\text{ep},J-1} C_{J-1}(t),$$
$$C_{\text{measured}}(t) = C_1(t) + \cdots + C_{J-1}(t) + K_p C_p(t), \quad (1)$$

where $C_j(t)$ is the tracer concentrations in the interstitial space weighted by the fractional interstitial volume in tissue-type $j$ at time $t$ for $j=1,\ldots,J$, where $J$ is the total number of vascular compartments; $C_p(t)$ is the tracer concentration in plasma (tracer input function); $C_{\text{measured}}(t)$ is the measured tracer concentration; $K_j^{\text{trans}}$ is the unidirectional volume transfer constant (/min) from plasma to tissue-type $j$; $k_{\text{ep},j}$ is the flux rate constants (/min) in tissue-type $j$; and $K_p$ is the plasma volume[27].

Solving (1) leads to $C_j(t) = C_p(t) \otimes \exp(-k_{\text{ep},j} t)$, $j=1, \ldots, J-1$, where $\otimes$ denotes the mathematical convolution, and $C_J(t) = C_p(t)$. The spatial-temporal patterns of tracer concentrations (pixel time-course) can be expressed as[28]

$$\begin{bmatrix} C_{\text{measured}}(i,t_1) \\ C_{\text{measured}}(i,t_2) \\ \vdots \\ C_{\text{measured}}(i,t_L) \end{bmatrix} = \begin{bmatrix} C_1(t_1) & \cdots & C_{J-1}(t_1) & C_p(t_1) \\ C_1(t_2) & \cdots & C_{J-1}(t_2) & C_p(t_2) \\ \vdots & \cdots & \vdots & \vdots \\ C_1(t_L) & \cdots & C_{J-1}(t_L) & C_p(t_L) \end{bmatrix} \begin{bmatrix} K_1^{\text{trans}}(i) \\ \vdots \\ K_{J-1}^{\text{trans}}(i) \\ K_p(i) \end{bmatrix}, \quad (2)$$

where $C_{\text{measured}}(i,t_l)$ is the tracer concentration at time $t_l$ at pixel $i$, $L$ is the number of sampled time points, $K_1^{\text{trans}}(i),\ldots,K_{J-1}^{\text{trans}}(i)$ are the local volume transfer constants of the tissue-types 1 to (J-1), at pixel $i$, respectively; and $K_p(i)$ is the local plasma volume at pixel $i$.

**Parallelism between multi-tissue compartment modeling and the theory of convex sets.** Apply a sum-based normalization to pixel time-course $C_{\text{measured}}(i,t_l)$ and using vector-matrix notation, we can re-express (2) as

$$\mathbf{C}_{\text{measured}}(i) = \sum_{j=1}^{J} K_j^{\text{trans}}(i) \mathbf{C}_j, \quad (3)$$

where $K_j^{\text{trans}}(i)$ is accordingly normalized over $j$, $\mathbf{C}_{\text{measured}}(i)$ and $\mathbf{C}_j$ are the vector notations (over time) of pixel time course $C_{\text{measured}}(i,t_l)$ and compartment time course $C_j(t_l)$, respectively. Since $K_j^{\text{trans}}(i)$ is always non-negative, as a non-negative linear combination of $\mathcal{C}_J = \{\mathbf{C}_j\}$, the set of pixel time-course $\mathcal{C}_{measured} = \{\mathbf{C}_{measured}(i)\}$ forms a subset of the *convex set* readily defined by the set of $\{\mathbf{C}_j\}$

$$\mathcal{C}_{measured} = \left\{ \sum_{j=1}^{J} K_j^{trans}(i) \mathbf{C}_j, \ K_j^{trans}(i) \geq 0, \ \sum_{j=1}^{J} K_j^{trans}(i) = 1, \ i = 1, ..., N \right\}. \tag{4}$$

MTCM exploits the strong parallelism between the multi-compartment model (3) and the theory of convex set. The fundamental principle is that *latent* compartments are pure vasculatures in a tumor whose pharmacokinetics $\{\mathbf{C}_j\}$ reside at the extremities of the scatter simplex occupied by the pixel time-courses, and accordingly, the interior of the simplex is occupied by the partial-volume pixels (linear non-negative mixtures of compartments) (**Fig. 1b**). Estimates of compartment pharmacokinetics may then be derived from the vertices of the multifaceted simplex that most tightly encloses the pixel time-courses and has the same number of compartments as vertices (**Fig. 1d**) [29]. MTCM algorithm is supported theoretically by a well-grounded mathematical frameworkas summarised below (see formal proofs in **Appendix 2. Supplementary Method**).

**Theorem 1 (Convexity of pixel time-course).** *Suppose that the J compartment pharmacokinetics $\{\mathbf{C}_j\}$ are linearly independent, and $\mathbf{C}_{measured}(i) = \sum_{j=1}^{J} K_j^{trans}(i) \mathbf{C}_j$ where local volume transfer constants $\{K_j^{trans}(i)\}$ are non-negative and have at least one pixel whose signal is highly and exclusively enriched in a particular vascular compartment, then, $\mathcal{C}_{measured}$ uniquely specifies a convex set $\mathcal{H}\{\mathcal{C}_{measured}\} = \left\{ \sum_{i=1}^{N} \alpha_i \mathbf{C}_{measured}(i), \ \alpha_i \geq 0, \ \sum_{i=1}^{N} \alpha_i = 1 \right\}$ which is in fact identical to the convex set $\mathcal{H}\{\mathcal{C}_J\} = \left\{ \sum_{j=1}^{J} \beta_j \mathbf{C}_j, \ \beta_j \geq 0, \ \sum_{j=1}^{J} \beta_j = 1 \right\}$, whose vertices are the J compartment pharmacokinetics $\{\mathbf{C}_j\}$.*

**Theorem 2 (Optimum source dominance).** *Suppose that the local volume transfer constants $\{\mathbf{K}^{trans}(v_j) = [K_1^{trans}(v_j), ..., K_m^{trans}(v_j), ..., K_J^{trans}(v_j)]\}$ are the vertices of the pixel time-course scatter simplex. Then the MTCM solution based on these vertices $\{\mathbf{K}^{trans}(v_j)\}$ achieves the maximum source dominance in the sense of $K_m^{trans}(v_j) = \max_{i=1,2,...N} K_m^{trans}(i)$.*

From Theorems 1 and 2, there is a mathematical solution uniquely identifying the compartment model (3) based on the observed set of pixel time-course $\mathcal{C}_{measured} = \{\mathbf{C}_{measured}(i)\}$ (identifiability and optimality): The *vertices* of convex set $\mathcal{H}\{\mathcal{C}_{measured}\}$ are the $J$ compartment pharmacokinetics $\{\mathbf{C}_j\}$ when there is at least one pixel whose signal is highly and exclusively enriched in a particular vascular compartment(**Fig. 1b**). This means that, in principle, under a noise-free scenario, we can directly estimate $\{\mathbf{C}_j\}$ by locating the *vertices* of $\mathcal{H}\{\mathcal{C}_{measured}\}$ (**Fig. 1d**).

**Data preprocessing (Fig. 1a).** First, the tumor area is extracted by masking out the non-tumor tissues surrounding the tumor site [30]. Second, the first few image frames, such as the four initial images of DCE-MRI sequences in our experiments, are removed because they correspond to the time prior to sufficient onsite tracer uptake. Third, pixels whose temporal average signal intensity is lower than 5% of the maximum value, or whose temporal dynamic variation is lower than 5% of the maximum value, are eliminated, because these non-informative pixels could have a negative impact on subsequent analyses. Fourth, the pixel time series is normalized over time using a sum-based normalization scheme, focusing the analysis on the "shape" of pharmacokinetics rather than on absolute tracer concentration.

It is true that accurate extraction of tumor region is critical to any image-based analysis that is focused on tumor characterization, where non-tumor tissue would constitute a confounding factor. Theoretically, MTCM method can handle well such situation since it is a completely unsupervised approach, relying on the MDL-based model selection. Specifically, since MTCM is specifically designed to work on multiple tissue compartment modeling, when a significant portion of the surrounding healthy tissue is included in the processed 'tumor' area, the healthy tissue will be considered as an additional/individual compartment in Eq. (1) and Fig. 1c. The MDL-based model selection procedure will statistically determine the number of underlying tissue compartments in the processed area, e.g., whether the contribution of surrounding healthy tissues is significant to be considered as an independent compartment. Though MTCM methodology can accept the processed area extracted by any image segmentation methods, the tumor region in our study can be outlined by a collaborative effort by computer scientists and clinicians (Wang, Adali et al. 1998, Xuan, Adali et al. 2000, Li, Wang et al. 2001). In the case of heavy noise and fuzzy boundary, a consensus approach may be adopted that surveys the results of multiple methods.

**Clustering of pixel time-course (Fig. 1b).** To reduce the impact of noise/outlier data points, improve the efficiency of subsequent convex analysis of mixtures, and permit an automated determination of the number of underlying vascular compartments using the minimum description length (MDL) criterion, we aggregated pixel time-courses into representative clusters using a combined affinity propagation and expectation-maximization clustering [31] (**Appendix 2. Supplementary Method**).As an initialization-free and near-global-optimum clustering method, affinity propagation clustering (APC) simultaneously considers all data points as potential exemplars and recursively exchange real-valued messages between data points until a high-quality set of exemplars and corresponding clusters gradually emerges. Let the "similarity" $s(i,m)$ indicate how well the $m$th data point is suited to be the exemplar for $i$th data point; the "responsibility" $r(i,m)$ reflects the accumulated evidence for how well-suited the $m$th data point is to serve as the exemplar for the $i$th data point; the "availability" $a(i,m)$ reflects the accumulated evidence for how appropriate the $i$th data point chooses $m$th data point as its exemplar. Then, supposing that there are $N$ data points (e.g., pixels) in total, the responsibilities $r(i,m)$ are computed based on

$$r(i,m) \leftarrow s(i,m) - \max_{m' \in \{1, ..., N\} \cap m' \neq m} \{a(i,m') + s(i,m')\}, \quad (5)$$

where the availabilities $a(i,m)$ are initialized to zero and the competitive update rule (5) is purely data-driven. Whereas the responsibility update (5) allows all candidate exemplars to compete for ownership of a data point, the availability update rule

$$a(i,m) \leftarrow \min\left\{0, r(m,m) + \sum_{i' \in \{1, ..., N\}, i' \neq i, i' \neq m} \max\{0, r(i',m)\}\right\} \quad (6)$$

collects evidence from data points to support a good exemplar, where the "self-availability" is updated differently $a(m,m) \leftarrow \sum_{i' \notin m} \max\{0, r(i',m)\}$. Then, the availabilities and responsibilities are combined to identify exemplars $m^* = \arg\max_m \{a(i,m) + r(i,m)\}$. The update rules are repeated iteratively and terminated when no change occurs for 10 iterations [31].

**Convex analysis of mixtures (Fig. 1d).** To identify the *vertices* of convex set $\mathcal{H}\{\mathcal{C}_{\text{measured}}\}$, we performed convex analysis of mixtures (CAM) on the obtained $M$ cluster centers $\{\mathbf{C}_m\}$. We assumed $J$ vascular compartments and conducted an exhaustive combinatorial search (with total $C_J^M$ combinations), based on a convex-hull-to-data fitting criterion, to identify the most probable $J$ vertices (**Appendix 2. Supplementary Method**). We used the margin-of-error

$$\delta_{m,\{1,...J\} \in C_J^M} = \min_{\alpha_1,...\alpha_J} \left\| \mathbf{C}_m - \sum_{j=1}^{J} \alpha_j \mathbf{C}_j \right\|_2, \ \alpha_j \geq 0, \ \sum_{j=1}^{J} \alpha_j = 1, \qquad (7)$$

to quantify the distance between $\mathbf{C}_m$ and convex set $\mathcal{H}\{\mathcal{C}_{J \in C_J^M}\}$, where we have $\delta_{m,\{1,...J\} \in C_J^M} = 0$ if $\mathbf{C}_m$ is inside $\mathcal{H}\{\mathcal{C}_{J \in C_J^M}\}$. We then selected the most probable $J$ vertices when the corresponding sum of the margin between the convex hull and the remaining "exterior" cluster centers reaches its minimum:

$$\{1^*,...J^*\} = \arg\min_{\{1,...J\} \in C_J^M} \sum_{m=1}^{M} \delta_{m,\{1,...J\} \in C_J^M}. \qquad (8)$$

**Model selection procedure**. One important discovery step concerning MTCM is to detect the number $J$ of the underlying tissue compartments. We used MDL, a widely-adopted and consistent information theoretic criterion, to guide model selection (**Appendix 2. Supplementary Method**). We performed CAM on several competing candidates, and selected the optimal model that assigns high probabilities to the observed data while at the same time whose parameters are not too complex to encode [32]. Specifically, a model is selected with $J$ tissue compartments by minimizing the total description code length defined by

$$\text{MDL}(J) = -\log\left(\mathcal{L}\left(\mathcal{C}_M \mid \Theta(J)\right)\right) + \frac{J-1}{2}\log(M) + \frac{JM}{2}\log(L), \qquad (9)$$

where $\mathcal{L}(\cdot)$ denotes the joint likelihood function of the clustered compartment model, $\mathcal{C}_M$ denotes the set of $M$ cluster centers, and $\Theta(J)$ denotes the set of freely adjustable parameters in the clustered compartment model (see Supplementary Method).

**Estimation of pharmacokinetics parameters in MTCM.** Having determined the pure-volume pixels associated with the vertices of $\mathcal{H}\{\mathcal{C}_J\}$ and the value of $J$, we estimated the values of tissue-specific vascular compartment pharmacokinetics parameters, i.e., flux rate constants $\{k_{\text{ep},j}\}$ and volume transfer constants $\{K_j^{\text{trans}}\}$, where the vertex of fastest tracer enhancement (reaching its peak most rapidly) is identified as $\mathbf{C}_p$ (**Appendix 2.**

**Supplementary Method**). We constructed the Toeplitz matrix $\mathbf{H}(k_{ep,j})$ (sampled system impulse response) and solved the following optimization problem

$$\left\{\hat{k}_{ep,j}, \hat{K}_j^{trans}\right\} = \underset{K_j^{trans}, k_{ep,j}}{\arg\min} \left\| \mathbf{C}_{measured,j} - K_j^{trans} \mathbf{H}(k_{ep,j}) \mathbf{C}_p \right\|_2, \quad (10)$$

$$\text{subject to } K_j^{trans} > 0 \text{ and } k_{ep,j} > 0$$

for $j = 1, ..., J-1$. Subsequently, we estimated local volume transfer constants by solving

$$\left\{\hat{K}_j^{trans}(i)\right\} = \underset{\{K_j^{trans}(i)\}}{\arg\min} \left\| \mathbf{C}_{measured}(i) - \sum_{j=1}^{J} K_j^{trans}(i) \mathbf{C}_j(\hat{k}_{ep,j}) \right\|_2, \quad (11)$$

$$\text{subject to } K_j^{trans}(i) \geq 0, \ \forall j$$

which readily reveals the intratumor vascular heterogeneity.

**Synthetic DCE-MRI datasets.** We first validated MTCM-generated estimates of tissue-specific vascular pharmacokinetics parameters using a set of realistic synthetic DCE-MRI data with known parameter values. We set $J=3$, indexing two tissue compartments and one plasma input. We generated a large number of synthetic DCE-MRI time series by multiplying the customized local volume transfer constant maps $\{K_j^{trans}(i)\}$ with known compartment pharmacokinetics $\{\mathbf{C}_j(k_{ep,j})\}$. Synthetic data were comprised of 50 replicated datasets generated for each of the 12 parameter settings (**Supplementary Data 1**). We performed MTCM on all the datasets and compared the estimates of tissue-specific kinetic parameters produced by MTCM with the ground truth, in terms of both biases (accuracy) and variance (reproducibility) of the estimates, measured over 50 replicated datasets. For comparison purposes, we also evaluated the three most relevant methods (**Supplementary Table 1**). To determine whether the proposed MDL criterion detects the correct number of underlying tissue compartments, we calculated the MDL values for $J = 2, 3, ..., 6$ and identified the most probable value of $J$ when MDL achieves its minimum value(s) (**Fig. 3**).

**Characterization of differential vascular pharmacokinetics in advanced breast cancer case.** In the second application, we analyzed the real DCE-MRI data of an advanced breast tumor using both MTCM and the classic method. The T1-weighted gadolinium-enhanced (Gd-DTPA)DCE-MRI data set was acquired by three-dimensional scans performed every 30 seconds for a total of 11 minutes after the injection, on a 1.5 Tesla magnet using three-dimensional spoiled gradient-echo sequences (TR < 7 msec, TE < 1.5 msec, flip angle = 30°,

matrix = 192 × 256, 0.5 averages). Typically, 12-15 slices are obtained and 15-18 time frames are acquired for each case. We visually examined the convexity of projected pixel time-course via the top two convexity-preserved projections where the margin between the "exterior" data points and the convex hull is minimized. We observed that two-tissue compartments (a three-vertex convex set) were sufficient to describe the observed pixel time-course scatter simplex. While additional compartments can be used to account for outlier vertices, these compartments become difficult to interpret. We analyzed the dataset by setting $J = 4, 5, 6$ and observed noise-like and biologically implausible pharmacokinetics patterns. The minimum value of MDL confirmed $J$=3. The number of clusters $M$ takes values between 12 and 18, determined automatically by the APC algorithm.

**Characterizing longitudinal changes of differential vascular pharmacokinetics in treating angiogenic-active breast cancer case.** Vascular pharmacokinetics parameter values estimated by MTCM reveal longitudinal changes that may serve as the evidence of differential and heterogeneous responses to therapy. We analyzed the data sets arising from a longitudinal study of tumor response to anti-angiogenic therapy using similar imaging protocols (**Supplementary Data 3**). Three sets of DCE-MRI data were acquired during standard treatment, each three months apart, serving as the potential endpoints in assessing the response to therapy. To detect various yet potentially hidden changes accounting for intratumor heterogeneity, we applied the same MTCM and MDL (as well as the classic method) to the three data sets separately.

**Open source multiplatform standalone MTCM Java-R software.** Java GUI supported MTCM was implemented in both R and MATLAB, and runs on both Microsoft Windows and Linux platforms (http://mloss.org/software/view/437/). MTCM takes input the .mat data files that record the pixel time-course of DCE-MRI data in matrices. Each row corresponds to a time frame and each column corresponds to a pixel. Results of MTCM are provided to the users via a Java-based GUI (**Supplementary Fig. 5**).

## Appendix 1. Supplementary Discussion
**Dynamic intratumor heterogeneity: clonal repopulation & multi-compartment model.**
For the characterization of complex phenotypes and therapeutic responses [5,7,11], a major outstanding issue is how to accurately quantify intratumor vascular heterogeneity that may be severely confounded by the varying partial-volume effect [7,16]. Specifically, to capture the changes (that may reflect the underlying 'clonal' repopulation dynamics [34]) in (1) local volume transfer maps, (2) compartment pharmacokinetics, and (3) number of distinctive compartments, a completely unsupervised learning method is required to solve the multi-compartment model based on only observed DCE-MRI data.

MTCM using DCE-MRI has the potential to reveal functional intratumor vascular heterogeneity, without any type of external information, thus is an unbiased and data-driven approach. This advantage has significant implications. The recent results of Kreso *et al*. strongly suggest that biological differences between tumor cells can be due to additional mechanisms, other than genetic heterogeneities [34]. It has been reported, despite validated genetic homogeneity, the different cancer 'clones' observed in a tumor, displayed notable differences in behavior during the experiments [34]. More interestingly, some of these unusual clones remained inactive initially but reemerged at later stage; therapeutic drugs preferentially eliminated persistent clones while increased the proportion of clones that were initially dormant [34]. A more likely explanation about this clonal repopulation is the involvement of one or more distinct semistable epigenetic states, on which MTCM can help to reveal quantitatively at phenotypic and functional level. To our best knowledge, MTCM is the first tool of its type to address the suggestion that "Improved mathematical models built on actual clinical and experimental observations, will likely allow us to construct these pictures in the not-so-distant future [34]." For example, the results of MTCM may provide mechanistic models of tumor progression and responses to refine and optimize biopsy sampling, in terms of timing, number, and location of biopsies. Furthermore, some of form of intratumor heterogeneity (ITH) index may be subsequently defined as a predictor of tumor history and behaviour [38].

In complex tissues, functional heterogeneity is of great interest since it represents the integration of various upstream factors. Since cell-cell signalling plays a critical role in tumor development and evolution, cellular heterogeneity may constitutes only a partial picture, and dissection based on cell types may provide limited information since functional aspect is largely missing [39]. MTCM (a completely unsupervised method empowered by MDL based model selection) and DCE-MRI provide a powerful and in vivo method to reveal and quantify functional heterogeneity. In contrast, many conventional analytic methods (relying on prior knowledge) may miss the low-frequency yet critical subclone(s).

**Detecting the number of tissue compartment (model selection).**
To discover and characterize intratumor vascular heterogeneity, the true number of the underlying tissue compartments is an unknown 'structural' model parameter and must be estimated from the data [33,34]. To assure that MTCM provides a completely unsupervised machine learning method, we exploit the information theoretical criterion called minimum description length (MDL) [32,35], to detect the most appropriate number of tissue compartments in our analysis. From the MDL principle, we derived the specific MDL for MTCM model as Eq. (9) (see Supplementary Method).

MDL calculates the total number of 'bits' that are required to encode/explain both the 'data' and 'model'. When the model is given (or estimated), only information about 'mismatch' between model and data needs to be explained (or encoded). The first term (negative joint likelihood) in the MDL determines exactly the 'bits' for explaining the 'data' conditioned on the given model. The second and third terms represents the 'penalty' on the model complexity, that is, the total number of bits for explaining the model. Each of these terms involves two multiplicative factors: the number of free-adjustable parameters and the original data points used to estimate the parameter (or the original data points the parameter is used to or can 'explain'). Specifically, when estimating the compartment TCs (the column vectors of mixing matrix) $C_J = \{\mathbf{C}_1,...,\mathbf{C}_J\}$ parameterized by $J-1$ independent $k_{ep}$, we use some form of vector-average operation (i.e., $\min_{\alpha_{m_1}...\alpha_{m_J}} \left\| \boldsymbol{\mu}_{\mathbf{C}_{measured},m} - \sum_{j=1}^{J} \alpha_{m_j} \boldsymbol{\mu}_{\mathbf{C}_{measured},m_j} \right\|_2$; where nonnegative $\alpha_{m_1},...,\alpha_{m_J}$, with sum equal to 1, are the coefficients for defining a convex hull), the scalar entry $C_j(t_l)$ in $\mathbf{C}_j$ is estimated involving only $M$ scalar entries $\mu_{\mathbf{C}_{measured}}(m,t_l)$ for a given $t_l$, contributing total $(J-1)\log(M)/2$ bits. Similarly, when estimating the local volume transfer constants (the row vectors of sources) $\mathbf{K}^{trans}(m)$ with total $JM$ entries, we use some of form of vector-average operation (i.e., solving linear equations), where the scalar entry $K_j^{trans}(m)$ in $\mathbf{K}^{trans}(m)$ is estimated involving only $L$ scalar entries $\mu_{\mathbf{C}_{measured}}(m,t_l)$ for a given $m$, contributing total $JM\log(L)/2$ bits.

**Data quality control (QC).**

Quality control should be applied to reduce error in all image parameters. The impact of motion should be assessed and tumours for which parameter estimates are unreliable should be rejected. The level of bulk motion can be assessed for each tumor by first extracting an averaged time series plot for each tumor region of interest (ROI) on each slice in the imaging volume and then by visual assessment of the dynamic time series images. In- and through-plane motion can be investigated and a categorical score can be assigned for each tumor based on the evaluations of bulk motion. Tumors with a motion assessment score higher than a pre-specified threshold should be excluded [38].

## Appendix 2. Supplementary Method

**Parallelism between multi-compartment modeling and the theory of convex sets.**

We now discuss the identifiability of the compartment model (2) and the required conditions via the following definitions and theorems.

***Definition* 1.** Given a set of compartment TCs $\mathcal{C}_J = \{\mathbf{C}_1,...,\mathbf{C}_J\}$, we denote the convex set it specifies by

$$\mathcal{H}\{\mathcal{C}_J\} = \left\{\sum_{j=1}^{J} \alpha_j \mathbf{C}_j \mid \mathbf{C}_j \in \mathcal{C}_J,\ \alpha_j \geq 0,\ \sum_{j=1}^{J} \alpha_j = 1\right\}. \tag{S1}$$

***Definition* 2.** A compartment TC vector $\mathbf{C}_j$ is a vertex point of the convex set $\mathcal{H}\{\mathcal{C}_J\}$ if it can only be expressed as a trivial convex combination of $\mathbf{C}_1,...,\mathbf{C}_J$.

**Lemma 1 (Convex envelope of pixel time series).** *Suppose that the J compartment pharmacokinetics $\{\mathbf{C}_j\}$ are linearly independent, and $\mathbf{C}_{\text{measured}}(i) = \sum_{j=1}^{J} K_j^{\text{trans}}(i)\mathbf{C}_j$ where spatially-distributed local volume transfer constants $\{K_j^{\text{trans}}(i)\}$ are non-negative and normalized. Then, the elements of $\mathcal{C}_{\text{measured}}$ (the pixel time series) are confined within a convex set $\mathcal{H}\{\mathcal{C}_J\}$ whose vertices are the J compartment TCs $\mathbf{C}_1,...,\mathbf{C}_J$.*

***Definition* 3.** Any pixel whose associated normalized spatially-distributed volume transfer constants are in the form of $\mathbf{K}^{\text{trans}}(i_{\text{WGP}(j)}) \approx \mathbf{e}_j$ is called a well-grounded point (WGP) and corresponds to a pure-volume pixel, where $\{\mathbf{e}_j\}$ is the standard basis of *J*-dimensional real space (the axes of the first quadrant). In other words, we define pure volume pixels (or well-grounded pixels) as the pixels that are occupied by only a single compartment tissue type.

*Proof of Lemma 1.* By the definition of convex set [43], the fact that, $\forall \{i,j\}\ K_j^{\text{trans}}(i) \geq 0$, $\sum_{j=1}^{J} K_j^{\text{trans}}(i) = 1$ and $\mathbf{C}_{\text{measured}}(i) = \sum_{j=1}^{J} K_j^{\text{trans}}(i)\mathbf{C}_j$ readily yield $\mathbf{C}_{\text{measured}}(i) \in \mathcal{H}\{\mathcal{C}_J\}$ where

$$\mathcal{H}\{\mathcal{C}_J\} = \left\{\sum\nolimits_{j=1}^{J} \alpha_j \mathbf{C}_j \mid \mathbf{C}_j \in \mathcal{C}_J, \ \alpha_j \geq 0, \ \sum\nolimits_{j=1}^{J} \alpha_j = 1\right\}. \tag{S2}$$

Since $\mathbf{C}_1,...,\mathbf{C}_J$ are linearly independent, it follows that

$$\sum\nolimits_{j=1}^{J} \alpha_j \mathbf{C}_j = \mathbf{0} \text{ iff } \alpha_j = 0 \ \forall j \tag{S3}$$

that also implies that $\forall j$

$$\mathbf{C}_j = \sum\nolimits_{j=1}^{J} \alpha'_j \mathbf{C}_j \text{ iff } [\alpha'_1,...,\alpha'_J]^T = \mathbf{e}_j \ \forall j \tag{S4}$$

*i.e.*, $\mathbf{C}_j$ can only be a trivial convex combination of $\mathbf{C}_1,...,\mathbf{C}_J$. Hence, by Definition 2,

$\mathbf{C}_1,...,\mathbf{C}_J$ are therefore the vertices of convex set $\mathcal{H}\{\mathcal{C}_J\}$.

*Proof of Theorem 1.* Since $\exists i_{\text{WGP}(j)}$, $\mathbf{K}^{\text{trans}}(i_{\text{WGP}(j)}) = \mathbf{e}_j \ \forall j$, and $\mathbf{C}_{\text{measured}}(i) = \sum\nolimits_{j=1}^{J} K_j^{\text{trans}}(i)\mathbf{C}_j$,

we have

$$\mathbf{C}_{\text{measured}}(i_{\text{WGP}(j)}) = \mathbf{C}_j. \tag{S5}$$

Then, for any $\mathbf{z} \in \mathcal{H}\{\mathcal{C}_J\} = \left\{\sum\nolimits_{j=1}^{J} \alpha_j \mathbf{C}_j \mid \mathbf{C}_j \in \mathcal{C}_J, \ \alpha_j \geq 0, \ \sum\nolimits_{j=1}^{J} \alpha_j = 1\right\}$, we have

$$\begin{aligned}
\mathbf{z} &= \sum\nolimits_{j=1}^{J} \alpha_j \mathbf{C}_j \\
&= \sum\nolimits_{j=1}^{J} \alpha_j \mathbf{C}_{\text{measured}}(i_{\text{WGP}(j)}) \\
&= \sum\nolimits_{i=1}^{N} \alpha'_i \mathbf{C}_{\text{measured}}(i), \text{ where } \alpha'_i = \begin{cases} \alpha_j, & i \in \{i_{\text{WGP}(j)}\}, \\ 0, & i \notin \{i_{\text{WGP}(j)}\}, \end{cases}
\end{aligned} \tag{S6}$$

that implies $\mathbf{z} \in \mathcal{H}\{\mathcal{C}_{\text{measured}}\} = \left\{\sum\nolimits_{i=1}^{N} \alpha'_i \mathbf{C}_{\text{measured}}(i) \mid \mathbf{C}_{\text{measured}}(i) \in \mathcal{C}_{\text{measured}}, \ \alpha'_i \geq 0, \ \sum\nolimits_{i=1}^{N} \alpha'_i = 1\right\}$, *i.e.*,

$\mathcal{H}\{\mathcal{C}_J\} \subseteq \mathcal{H}\{\mathcal{C}_{\text{measured}}\}$. On the other hand, for any

$\mathbf{z} \in \mathcal{H}\{\mathcal{C}_{\text{measured}}\} = \left\{\sum\nolimits_{i=1}^{N} \alpha'_i \mathbf{C}_{\text{measured}}(i) \mid \mathbf{C}_{\text{measured}}(i) \in \mathcal{C}_{\text{measured}}, \ \alpha'_i \geq 0, \ \sum\nolimits_{i=1}^{N} \alpha'_i = 1\right\}$, we have

$$\begin{aligned}
\mathbf{z} &= \sum\nolimits_{i=1}^{N} \alpha_i \mathbf{C}_{\text{measured}}(i) \\
&= \sum\nolimits_{i=1}^{N} \alpha_i \sum\nolimits_{j=1}^{J} K_j^{\text{trans}}(i) \mathbf{C}_j \\
&= \sum\nolimits_{j=1}^{J} \left[\sum\nolimits_{i=1}^{N} \alpha_i K_j^{\text{trans}}(i)\right] \mathbf{C}_j \\
&= \sum\nolimits_{j=1}^{J} \beta_j \mathbf{C}_j, \text{ where } \beta_j = \sum\nolimits_{i=1}^{N} \alpha_i K_j^{\text{trans}}(i) \text{ and } \sum\nolimits_{j=1}^{J} \beta_j = 1,
\end{aligned} \tag{S7}$$

that implies $\mathbf{z} \in \mathcal{H}\{\mathcal{C}_J\} = \left\{\sum\nolimits_{j=1}^{J} \alpha_j \mathbf{C}_j \mid \mathbf{C}_j \in \mathcal{C}_J, \ \alpha_j \geq 0, \ \sum\nolimits_{j=1}^{J} \alpha_j = 1\right\}$, *i.e.*, $\mathcal{H}\{\mathcal{C}_{\text{measured}}\} \subseteq \mathcal{H}\{\mathcal{C}_J\}$.

Combining $\mathcal{H}\{\mathcal{C}_J\} \subseteq \mathcal{H}\{\mathcal{C}_{\text{measured}}\}$ and $\mathcal{H}\{\mathcal{C}_{\text{measured}}\} \subseteq \mathcal{H}\{\mathcal{C}_J\}$ gives $\mathcal{H}\{\mathcal{C}_{\text{measured}}\} = \mathcal{H}\{\mathcal{C}_J\}$, and together with Lemma 1 readily completes the proof of Theorem 1.

*Proof of theorem 2.* Consider the pixel $\mathbf{K}^{\text{trans}}(i^*) = \sum_{j=1}^{J} \alpha_j(i^*) \mathbf{K}^{\text{trans}}(v_j)$ of the convex hull defined by the vertices $\{\mathbf{K}^{\text{trans}}(v_j)\}$ whose $m$th entry is the largest among all pixels, i.e., $K_m^{\text{trans}}(i^*) = \max_{i=1,2,...N} K_m^{\text{trans}}(i)$. Since $\sum_{j=1}^{J} \alpha_j(i^*) = 1$, we may therefore write

$$K_m^{\text{trans}}(i^*) = \left(\sum_{j=1}^{J} \alpha_j(i^*)\right) K_m^{\text{trans}}(i^*) = \sum_{j=1}^{J} \alpha_j(i^*) K_m^{\text{trans}}(i^*). \tag{S8}$$

Alternatively, the $m$th entry of $\mathbf{K}^{\text{trans}}(i^*)$ can be expressed as

$$K_m^{\text{trans}}(i^*) = \sum_{j=1}^{J} \alpha_j(i^*) K_m^{\text{trans}}(v_j). \tag{S9}$$

By the unique convex expression of $K_m^{\text{trans}}(i^*)$, we have

$$\sum_{j=1}^{J} \alpha_j(i^*) \left(K_m^{\text{trans}}(i^*) - K_m^{\text{trans}}(v_j)\right) = 0, \tag{S10}$$

which, together with the fact $\alpha_j(i^*) \geq 0$ and $K_m^{\text{trans}}(i^*) - K_m^{\text{trans}}(v_j) \geq 0$, implies $i^* \in \{v_j\}$.

**Clustering of Pixel Time Series**

The purpose of multivariate clustering of normalized pixel time series is three-fold: 1) data clustering has proven to be an effective tool for reducing the impact of noise/outlier data points on model learning [44-46]; 2) aggregation of pixel time series into a few clusters improves the efficiency of subsequent convex analysis of mixtures [36,37]; 3) the resultant clustered compartment model permits an automated determination of the number of underlying tissue compartments using the minimum description length (MDL) criterion [33,38,39].

There has been considerable success in using SFNMs to model clustered data sets such as DCE-MRI data, taking a sum of the following general form [40-42]:

$$p(\mathbf{K}^{\text{trans}}(i)) = \sum_{m=1}^{J} \pi_m g(\mathbf{K}^{\text{trans}}(i) | \mathbf{e}_m, \boldsymbol{\Sigma}_{\mathbf{K}^{\text{trans}},m}) + \sum_{m=J+1}^{M} \pi_m g(\mathbf{K}^{\text{trans}}(i) | \boldsymbol{\mu}_{\mathbf{K}^{\text{trans}},m}, \boldsymbol{\Sigma}_{\mathbf{K}^{\text{trans}},m}), \tag{S11}$$

where the first term corresponds to the clusters of pure volume pixels $(m = 1,...,J)$, the second term corresponds to the clusters of partial volume pixels $(m = J+1,...,M)$, $M$ is the total number of pixel clusters, $\pi_m$ is the mixing factor, $g(\cdot)$ is the Gaussian kernel, $\mathbf{e}_m$ denotes the $m$th natural basis vector corresponding to the mean vector of the $m$th pure tissue compartment, and $\boldsymbol{\mu}_{\mathbf{K}^{\text{trans}},m}$ and $\boldsymbol{\Sigma}_{\mathbf{K}^{\text{trans}},m}$ are the mean vector and covariance matrix of cluster $m$, respectively. It is worth noting that SFNMs are a flexible and powerful statistical modeling tool and can adequately model clustered structure with essentially arbitrary complexity by introducing a sufficient number of mixture components. Thus, strict adherence of the (in general, unknown) ground-truth data distribution to the form in (9) is not required in most real-world applications [40,41]. By incorporating (7) into (9), the SFNM model for pixel time series becomes:

$$p(\mathbf{C}_{\text{measured}}(i)) = \sum_{m=1}^{J} \pi_m g\left(\mathbf{C}_{\text{measured}}(i) \mid \mathbf{a}_m, \mathbf{\Sigma}_{\mathbf{C}_{\text{measured}},m}\right) + \sum_{m=J+1}^{M} \pi_m g\left(\mathbf{C}_{\text{measured}}(i) \mid \mathbf{\mu}_{\mathbf{C}_{\text{measured}},m}, \mathbf{\Sigma}_{\mathbf{C}_{\text{measured}},m}\right)$$

(S12)

where $\mathbf{\Sigma}_{\mathbf{C}_{\text{measured}},m} = \mathbf{A}\mathbf{\Sigma}_{\mathbf{K}^{\text{trans}},m}\mathbf{A}^T$ and $\mathbf{\mu}_{\mathbf{C}_{\text{measured}},m} = \mathbf{A}\mathbf{\mu}_{\mathbf{K}^{\text{trans}},m}$, with $\mathbf{C} = [\mathbf{C}_1,...,\mathbf{C}_J]$. Accordingly, the first term of (S12) represents the corner clusters and the second term of (S12) represents the interior clusters (as shown in Fig. 1). From Theorem 1 and SFNM model (S12), the clustered pixel time series set $\mathcal{C}_{\text{measured}}$ is (approximately) confined within a convex set whose corner centers are the $J$ compartment TCs $\mathbf{C}_1,...,\mathbf{C}_J$.

It has been shown that significant computational savings can be achieved by using the EM algorithm to allow a mixture of the form (S12) to be fitted to the data [40,41]. Determination of the parameters of the model (S12) can be viewed as a "missing data" problem in which the missing information corresponds to pixel labels $l_{im} = \mathbf{I}(i,m)$ specifying which cluster generated each data point with $\mathbf{I}(i,m)$ denoting the indicator function. When no information about $l_{im}$ is available, the log-likelihood for the model (10) takes the form

$$\log \mathcal{L}(\mathcal{C}_{\text{measured}} \mid \mathbf{\Theta}) = \sum_{i=1}^{N} \log \left\{ \sum_{m=1}^{M} \pi_m g\left[\mathbf{C}_{\text{measured}}(i) \mid \mathbf{\mu}_{\mathbf{C}_{\text{measured}},m}, \mathbf{\Sigma}_{\mathbf{C}_{\text{measured}},m}\right] \right\}, \quad (S13)$$

where $\mathcal{L}(\cdot)$ denotes the joint likelihood function of SFNM and $\mathbf{\Theta} = \{\pi_m, \mathbf{\mu}_{\mathbf{C}_{\text{measured}},m}, \mathbf{\Sigma}_{\mathbf{C}_{\text{measured}},m}, \forall m\}$. If, however, we were given a set of already clustered data with specified pixel labels, then the log likelihood (known as the "complete" data log-likelihood) becomes

$$\log \mathcal{L}(\mathcal{C}_{\text{measured}}, \mathbf{L} \mid \mathbf{\Theta}) = \sum_{i=1}^{N} \sum_{m=1}^{M} l_{im} \log \left\{ \pi_m g\left[\mathbf{C}_{\text{measured}}(i) \mid \mathbf{\mu}_{\mathbf{C}_{\text{measured}},m}, \mathbf{\Sigma}_{\mathbf{C}_{\text{measured}},m}\right] \right\}, \quad (S14)$$

where $\mathbf{L} = \{l_{im} \mid i = 1,...,N; m = 1,...,M\}$. Actually, we only have indirect, probabilistic, information in the form of the posterior responsibilities $z_{im}$ for each model $m$ having generated the pixel time series $\mathbf{C}_{\text{measured}}(i)$. Taking the expectation of (S14), we then obtain the complete data log likelihood in the form

$$\log \mathcal{L}(\mathcal{C}_{\text{measured}}, \mathbf{Z} \mid \boldsymbol{\Theta}) = \sum_{i=1}^{N} \sum_{m=1}^{M} z_{im} \log \left\{ \pi_m g \left[ \mathbf{C}_{\text{measured}}(i) \mid \boldsymbol{\mu}_{\mathbf{C}_{\text{measured}},m}, \boldsymbol{\Sigma}_{\mathbf{C}_{\text{measured}},m} \right] \right\}, \quad (S15)$$

in which the $z_{im} = \Pr(l_{im} = 1 \mid \mathbf{C}_{\text{measured}}(i))$ are constants, and $\mathbf{Z} = \{z_{im} \mid i = 1,...,N; m = 1,...,M\}$.

Maximization of (S15) can be performed using the two-stage form of the EM algorithm, where the pixel labels $l_{im}$ are treated as missing data as aforementioned. At each complete cycle of the algorithm we commence with an "old" set of parameter values $\boldsymbol{\Theta}$. We first use these parameters in the E-step to evaluate the posterior probabilities $z_{im}$ using Bayes theorem [43]

$$z_{im} = \Pr(l_{im} = 1 \mid \mathbf{C}_{\text{measured}}(i)) = \frac{\pi_m g \left[ \mathbf{C}_{\text{measured}}(i) \mid \boldsymbol{\mu}_{\mathbf{C}_{\text{measured}},m}, \boldsymbol{\Sigma}_{\mathbf{C}_{\text{measured}},m} \right]}{\sum_{m^*=1}^{M} \pi_{m^*} g \left[ \mathbf{C}_{\text{measured}}(i) \mid \boldsymbol{\mu}_{\mathbf{C}_{\text{measured}},m^*}, \boldsymbol{\Sigma}_{\mathbf{C}_{\text{measured}},m^*} \right]}, \quad m \in \{1,...,M\}.$$

(S16)

These posterior probabilities are then used in the M-step to obtain "new" values $\boldsymbol{\Theta}$ using the following re-estimation formulas

$$\pi_m = \frac{1}{N} \sum_{i=1}^{N} z_{im}, \quad (S17)$$

$$\boldsymbol{\mu}_{\mathbf{C}_{\text{measured}},m} = \frac{\sum_{i=1}^{N} z_{im} \mathbf{C}_{\text{measured}}(i)}{\sum_{i=1}^{N} z_{im}}, \quad (S18)$$

$$\boldsymbol{\Sigma}_{\mathbf{C}_{\text{measured}},m} = \frac{\sum_{i=1}^{N} z_{im} \left( \mathbf{C}_{\text{measured}}(i) - \boldsymbol{\mu}_{\mathbf{C}_{\text{measured}},m} \right) \left( \mathbf{C}_{\text{measured}}(i) - \boldsymbol{\mu}_{\mathbf{C}_{\text{measured}},m} \right)^T}{\sum_{i=1}^{N} z_{im}}. \quad (S19)$$

**Convex Analysis of Mixtures**

At this point in our analysis procedure, each pixel time-course is represented by a group of cluster centers, where both the dimensionality and noise/outlier effect are significantly reduced. Given the obtained $M$ cluster centers $\boldsymbol{\mu}_{\mathbf{C}_{\text{measured}},1},...,\boldsymbol{\mu}_{\mathbf{C}_{\text{measured}},M}$, CAM is applied to separate pure-volume clusters from partial-volume clusters by detecting the "corners" of the convex hull containing all clusters of pixel TCs (theoretically supported by Lemma 1 and Theorem 1). Assuming the number of compartments $J$ is known a priori, an exhaustive combinatorial search (with total $C_J^M$ combinations), based on a convex-hull-to-data fitting criterion, is performed to identify the most probable $J$ corners. This explicitly maps pure-volume pixels to the corners and partial-volume pixels to the interior clusters of the convex hull.

Let $\{\boldsymbol{\mu}_{C_{measured},m_1},...,\boldsymbol{\mu}_{C_{measured},m_J}\}$ be any size-$J$ subset of $\{\boldsymbol{\mu}_{C_{measured},1},...,\boldsymbol{\mu}_{C_{measured},M}\}$. Then, the margin (*i.e.*, distance) between $\boldsymbol{\mu}_{C_{measured},m}$ and the convex hull $\mathcal{H}\{\boldsymbol{\mu}_{C_{measured},m_1},...,\boldsymbol{\mu}_{C_{measured},m_J}\}$ is computed by

$$\delta_{m,(m_1,...,m_J)} = \min_{\alpha_{m_1},...\alpha_{m_J}} \left\| \boldsymbol{\mu}_{C_{measured},m} - \sum_{j=1}^{J} \alpha_{m_j} \boldsymbol{\mu}_{C_{measured},m_j} \right\|_2, \quad (S20)$$

where $\alpha_{m_j} \geq 0$, $\sum_{j=1}^{J} \alpha_{m_j} = 1$. It shall be noted that if $\boldsymbol{\mu}_{C_{measured},m}$ is inside $\mathcal{H}\{\boldsymbol{\mu}_{C_{measured},m_1},...,\boldsymbol{\mu}_{C_{measured},m_J}\}$ then $\delta_{m,(m_1,...,m_J)} = 0$. Next, we define the convex-hull-to-data fitting error as the sum of the margin between the convex hull and the "exterior" cluster centers and detect the most probably $J$ corners with cluster indices $(m_1^*,...m_J^*)$ when the criterion function reaches its minimum:

$$(m_1^*,...m_J^*) = \underset{(m_1,...,m_J)}{\arg\min} \sum_{m=1}^{M} \delta_{m,(m_1,...,m_J)}. \quad (S21)$$

The optimization problems of (22) and (23) can be solved by advanced convex optimization procedure described in [43] and an exhaustive combinatorial search (for realistic values of $J$ and $M$, in practice), respectively.

**Model selection procedure**.

One important issue concerning MTCM method is the detection of the structural parameter $J$ in the model (the number of underlying tissue compartments or types), often called model selection [32,44,45]. This is indeed particularly critical in real-world applications where the true structure of the compartment models may be unknown a priori. We propose to use a widely-adopted and consistent information theoretic criterion, namely the minimum description length (MDL) [32,39,44], to guide model selection. The major thrust of this approach is the formulation of a model fitting procedure in which an optimal model is selected from several competing candidates, such that the selected model best fits the observed data. MDL formulates the problem explicitly as an information coding problem in which the best model fit is measured such that it assigns high probabilities to the observed data while at the same time the model itself is not too complex to describe.

However, when the number of pixels is large as in DCE-MRI application, direct use of MDL may underestimate the value of $J$, due to the lack of "structure" in classical compartment models (over-parameterization) [33,38,39]. We therefore propose to naturally adopt and extend the clustered compartment models into the MDL formulation [52]. The proposed clustered compartment model allows all pixels belonging to the same cluster to share a common $\mathbf{K}_m^{\text{trans}}$, thus greatly reducing model complexity for a given value of $J$ (the number of convex hull corners). Specifically, a model is selected with $J$ tissue compartments by minimizing the total description length defined by [33,44]

$$\text{MDL}(J) = -\log\left(\mathcal{L}\left(\mathcal{C}_{\text{measured},M} \mid \Theta(J)\right)\right) + \frac{J-1}{2}\log(M) + \frac{JM}{2}\log(L), \quad (S22)$$

where $\mathcal{L}(\cdot)$ denotes the joint likelihood function of the clustered compartment model, $\mathcal{C}_{\text{measured},M}$ denotes the set of $M$ cluster centers, and $\Theta(J)$ denotes the set of freely adjustable parameters in the clustered compartment model.

Our aim herein is to use MDL criterion [33,44] and the MTCM estimates to select the best value of $J$ automatically (the number of convex hull corners or tissue compartments). In the clustered $J$-tissue compartment model, we allow all pixels belonging to the same cluster to share a common local volume transfer constant, namely $\mathbf{K}_m^{\text{trans}}$ with length $J$, $m=1,...,M$. Letting $\mathbf{C}_{\text{measured},m}$ be the $m$th cluster center associated with $\mathbf{K}_m^{\text{trans}}$, from (4), we can express the clustered compartment model as follows:

$$\mathbf{C}_{\text{measured},m} = [\mathbf{C}_1,...,\mathbf{C}_J]\mathbf{K}_m^{\text{trans}} + \mathbf{n}, \quad (S23)$$

where $\mathbf{n}$ is the modeling residual noise assumed to follow zero-mean white Gaussian distribution $\mathbf{n} \sim N(\mathbf{0},\sigma^2 \cdot \mathbf{I})$ with variance $\sigma^2$.

We specify $\Theta(J)$ as follows. From equations (3)-(4), $\mathbf{C}_1,...,\mathbf{C}_{J-1}$ are parameterized by $k_{\text{ep},j}, j=1,2,...,J-1$. Furthermore, $\mathbf{C}_p$ is parameterized by $\{\lambda_1,\lambda_2,\lambda_3,\alpha_2,\alpha_3\}$ based on the well-known exponential model [38,46]. Then, together with $\mathbf{K}_1^{\text{trans}},...,\mathbf{K}_M^{\text{trans}}$ and $\sigma$, we have $\Theta(J) = \{\lambda_1,\lambda_2,\lambda_3,\alpha_2,\alpha_3,k_{\text{ep},1},k_{\text{ep},2},...k_{\text{ep},J-1},\mathbf{K}_1^{\text{trans}},...,\mathbf{K}_M^{\text{trans}},\sigma\}$.

Based on MTCM estimated $\mathbf{C}_1,...,\mathbf{C}_J$ determined by $\hat{\mathbf{C}}_p$ and $\{\hat{k}_{\text{ep},j}, j=1,...,J-1\}$, and $\mathbf{K}_m^{\text{trans}}$ and $\sigma$ obtained by the maximum-likelihood estimation,

$$\hat{\mathbf{K}}_m^{\text{trans}} = \arg\max_{\mathbf{K}_m}\left(\log(\mathcal{L}(\mathcal{C}_{\text{measured},M} \mid \Theta(J)))\right) = \arg\min_{\mathbf{K}_m}\left(\|\mathbf{C}_{\text{measured},m} - [\mathbf{C}_1,...,\mathbf{C}_J]\mathbf{K}_m^{\text{trans}}\|^2\right),$$
$$\text{s.t. } \mathbf{K}_m^{\text{trans}} \geq 0 \quad \forall m$$

$$\hat{\sigma} = \arg\max_{\sigma}\left(\log(\mathcal{L}(\mathcal{C}_{\text{measured},M} \mid \Theta(J)))\right) = \sqrt{\frac{1}{M \cdot L}\sum_{m=1}^{M}\|\mathbf{C}_{\text{measured},m} - [\mathbf{C}_1,...,\mathbf{C}_J]\mathbf{K}_m^{\text{trans}}\|^2},$$

we can express the joint likelihood function in the MDL given by (S22) as

$$\mathcal{L}(\mathcal{C}_{\text{measured},M} \mid \Theta(J)) = \prod_{m=1}^{M} \left(2\pi\sigma^2\right)^{-L/2} \exp\left(-\frac{\|\mathbf{C}_{\text{measured},m} - [\mathbf{C}_1,...,\mathbf{C}_J]\mathbf{K}_m^{\text{trans}}\|^2}{2\sigma^2}\right). \quad (S24)$$

**Estimation of pharmacokinetics parameters in MTCM.**

Having determined the probabilistic pixel memberships associated with pure-volume compartments, $z_{ij}$ for $j=1,...,J$, $i=1,...,N$, we can then estimate the tissue-specific compartmental parameters, namely $K_j^{\text{trans}}$ and $k_{\text{ep},j}$, $j=1,...,J$, directly from DCE-MRI pixel time series $\mathbf{C}_{\text{measured}}(i)$, in which various compartment modeling techniques can be readily applied.

To specify which "exterior" cluster is associated with which compartment, we investigate the temporal enhancement patterns of the "exterior" cluster centers. As aforementioned, $C_p$ is associated with the cluster of the fastest enhancement (reaching its peak most rapidly); $C_j$ is associated with the cluster of $j$th tissue type. We then compute $\mathbf{C}_p$ and $\mathbf{C}_j$ via

$$\mathbf{C}_p = \frac{\sum_{i=1}^{N} z_{iJ}\mathbf{C}_{\text{measured}}(i)}{\sum_{i=1}^{N} z_{iJ}}, \quad \mathbf{C}_j = \frac{\sum_{i=1}^{N} z_{ij}\mathbf{C}_{\text{measured}}(i)}{\sum_{i=1}^{N} z_{ij}}, \quad j=1,...,J-1. \quad (S25)$$

We then recall the relationship $C_j(t) = K_j^{\text{trans}} C_p(t) \otimes \exp(-k_{\text{ep},j}t)$, and discretize the convolution (with discretization interval $\Delta t$ (/min)) to the following vector-matrix notation

$$\mathbf{C}_{\text{measured},j} = K_j^{\text{trans}}\mathbf{H}(k_{\text{ep},j})\mathbf{C}_p, \quad j=1,...,J-1 \quad (S26)$$

by constructing a Toeplitz matrix

$$\mathbf{H}(k_{\text{ep},j}) = \begin{bmatrix} e^{-k_{\text{ep},j}t_1} & 0 & 0 & ... & 0 \\ e^{-k_{\text{ep},j}t_2} & e^{-k_{\text{ep},j}t_1} & 0 & ... & 0 \\ ... & ... & ... & ... & ... \\ e^{-k_{\text{ep},j}t_L} & e^{-k_{\text{ep},j}t_{L-1}} & e^{-k_{\text{ep},j}t_{L-2}} & ... & e^{-k_{\text{ep},j}t_1} \end{bmatrix} \Delta t \quad (S27)$$

that is the sampled system impulse response. Then, the estimate of $k_{\text{ep},j}$ and $K_j^{\text{trans}}$ can be obtained by solving the following optimization problem

$$\{\hat{k}_{\text{ep},j}, \hat{K}_j^{\text{trans}}\} = \arg\min_{K_f^{\text{trans}}, k_{\text{ep},f}} \|\mathbf{C}_{\text{measured},j} - K_j^{\text{trans}}\mathbf{H}(k_{\text{ep},j})\mathbf{C}_p\|_2$$
$$\text{s.t. } K_j^{\text{trans}} > 0, k_{\text{ep},j} > 0 \quad (S28)$$

for $j = 1,...,J-1$. Finally, we can calculate the compartment TCs $\mathbf{C}(\mathbf{C}_p, \hat{k}_{ep,1},...,\hat{k}_{ep,J-1})$ based on $\mathbf{C}_p$ and $\hat{k}_{ep,1},...,\hat{k}_{ep,J-1}$, and then estimate the local volume transfer constants $\mathbf{K}^{trans}(i) = \left[K_1^{trans}(i),...,K_{J-1}^{trans}(i), K_p(i)\right]^T$ based on equation (4) via

$$\hat{\mathbf{K}}^{trans}(i) = \underset{\mathbf{K}^{trans}(i)}{\arg\min} \left\| \mathbf{C}_{measured}(i) - \mathbf{C}(\mathbf{C}_p, \hat{k}_{ep,1},...,\hat{k}_{ep,J-1}) \mathbf{K}^{trans}(i) \right\|_2 \quad (S29)$$
$$s.t. \ K_1^{trans}(i) \geq 0,..., K_{J-1}^{trans}(i) \geq 0, K_p(i) \geq 0,$$

that reflects the spatial heterogeneity of vascular permeability [53].

## Acknowledgement

This work was funded in part by the National Institutes of Health under Grants EB000830, NS29525, and CA149147. We thank T.H. Chan and D.J. Miller for technical assistance and discussion.

## Author Contributions

Y.W. and L.C. developed MTCM and wrote the manuscript; L.C. and N.W. applied MTCM and performed the experiments; P.L.C and E.M.C.H provided imaging expertise to the project and experiment; R.C. and Z.M.B provided biological expertise to the manuscript.

**Figure Legends**

Figure 1 | The proposed multi-tissue compartment modeling pipeline for uncovering intratumor vascular heterogeneity. (a) On the DCE-MRI sequence, tumor region is extracted using a digital mask. Then, pixel time-courses are collected and normalized over time. (b) Pixel time-courses are grouped into clusters with initialization-free multivariate clustering techniques. On the simplex of pixel time-courses, the clusters present at the vertices are identified by a convex analysis of mixtures. (c) Using pure-volume pixels, multi-compartment modeling is performed to estimate tissue-specific flux rate constants and volume transfer constants. (d) Scatter simplex of real DCE-MRI data from an advanced breast cancer. (e) Estimated tissue-specific compartmental time-activity curves: 'blue' – plasma input function; 'red' – fast flow kinetics; 'green' – slow flow kinetics; and example images of the associated local volume transfer constants. (f) Illustrative microscopic images of normal and abnormal vessel architecture (McDonald and Choyke, *Nat Med* **9**, 2003).

Figure 2 | Quantitative estimates of tissue-specific pharmacokinetic parameters in a longitudinal breast cancer study reveal changes in tumor vascular behavior in response to hybrid anti-angiogenesis chemotherapy. While tumor size regression (largely determined by bulk tumor populations rather than rarer cancer stem cells) is clearly observed, together with a transient "normalization", the detected tumor islands of persistent enhancement predict the confirmed recurrence despite the dramatic size changes . (a) Snapshots of DCE-MRI sequences taken from the same tumor before, during, and after therapy. (b) Scatter simplex of baseline DCE-MRI data taken before therapy; estimated tissue-specific compartmental time-activity curves; and example images of the associated local volume transfer constants. (c) Scatter simplex of interim DCE-MRI data taken during therapy; estimated tissue-specific compartmental time-activity curves; and example images of the associated local volume transfer constants. (d) Scatter simplex of closing DCE-MRI data taken after therapy; estimated tissue-specific compartmental time-activity curves; and example images of the associated local volume transfer constants.

Figure 3 | MTCM estimates time-activity curves in multiple vascular compartments simultaneously and quantitatively reconstructs tissue-specific local volume transfer constants - synthetic DCE-MRI experiments: (a) synthesis of image series; (b) scatter simplex of synthesized image series; (c) tissue-specific compartmental tracer concentration curves and local volume transfer constant maps, estimated by MTCM; (d) MDL model selection to detect the number of compartments; (e) tissue-specific compartmental tracer concentration curves estimated by principle component analysis (PCA).

**Support Information**

Supplementary Figure 1 – MTCM estimates time-activity curves in multiple vascular compartments simultaneously and quantitatively reconstructs tissue-specific local volume transfer constants – mouse DCE-MRI experimental data. (a) Snapshots of DCE-MRI sequence taken from the same tumor at 26 time points. Time point 1 is pre-contrast, and time points 2-26 are post-contrast. The first two time points are removed in the experiment. Each time point contains 4 sections from the same tumor. (b) The MDL curve of model selection and 3 is the optimal choice corresponding to the minimum MDL value. (c) Estimated tissue-specific compartmental time-activity curves: 'blue' - plasma input function; 'red' – fast flow kinetics; 'green' – slow flow kinetics. (d) Estimated maps of local volume transfer constants from four sections in the same tumor.

Supplementary Figure 2 – Comparison of time-activity curves of total vascular pool within the region of interests and tissue-specific time-activity curves estimated by MTCM, in a longitudinal DCE-MRI study on a breast cancer tumor: (a)-(c) time-activity curves of total vascular pool; (d)-(f) MTCM-estimated time-activity curves of fast flow pool; (g)-(i) MTCM-estimated time-activity curves of slow flow pool; (j)-(l) MTCM-estimated time-activity curves of plasma input function.

Supplementary Figure 3 – MTCM dissects tissue compartments into anatomical structures of the mouse using dynamic fluorescence molecular imaging data acquired on a mouse after bolus injection of indocyanine green dye, allowing the longitudinal identification of the internal organs. (a) Physiologically interpretable biodistribution dynamics of the major organs with ten fluorescence time courses showing distinct patterns of circulating, accumulating, or metabolizing the dye in different organs. (b) The merged and color-coded maps of the dissected tissue compartments agree well with a digital anatomical mouse atlas. (c) The gray-scale maps of the dissected individual tissue compartment (Kidney: $K^{trans}$=1.0004, $k_{ep}$=0.0134; Spine: $K^{trans}$=1.0269, $k_{ep}$=0.0241; Antipose: $K^{trans}$=0.7333, $k_{ep}$=0.0100; Large intestine: $K^{trans}$=0.7808, $k_{ep}$=0.0203; Nodes: $K^{trans}$=0.6719, $k_{ep}$=0.0049; Blood vessels: $K^{trans}$=0.9891, $k_{ep}$=0.0222; Liver: $K^{trans}$=0.7839, $k_{ep}$=0.0128; Brain: $K^{trans}$=0.7553, $k_{ep}$=0.0258; Stomach: $K^{trans}$=0.8955, $k_{ep}$=0.0143; Lung: $K^{trans}$=0.6656, $k_{ep}$=0.0167).

Supplementary Figure 4 – MTCM estimates time-activity curves in multiple DCE-MRI data produced in clinical practice. (a)-(c) show raw image series, scatter simplex of image series and estimated tissue-specific compartmental time-activity curves and local volume transfer constant maps, respectively for a case; (d)-(f) display the same things for another case.

Supplementary Figure 5 – MTCM software package in R and Java is developed to implement MTCM algorithm, as well as the other algorithms widely used in blind source separation. The user-friendly Java GUI (a) can generate the tissue-specific local volume transfer constants and pharmacokinetic parameters on the right. Two pop-up windows (b) will show the projection of clustered pixels on the simplex, and (c) will display the estimated tissue-specific compartmental time-activity curves.

Supplementary Table 1 – Comparison of tissue-specific kinetic parameter estimation by MTCM and three most relevant methods, based on synthetic DCE-MRI experimental data.

Supplementary Table 2 – Tissue-specific kinetic parameter estimates by MTCM on mouse DCE-MRI experimental data.

Supplementary Table 3 – MTCM estimates of flux rate constants and volume transfer constants of a breast cancer tumor before, during, and after treatment in the longitudinal study.

Supplementary Table 4 – Fractions of partial-volume pixels before, during, and after treatment in the longitudinal study.

Supplementary Table 5. MTCM estimated tissue heterogeneity score before, during, and after treatment in the longitudinal study.

Supplementary Data 1– Synthetic datasets generated for 12 parameter settings.

Supplementary Data 2 –DCE-MRI data sets arising from mouse DCE-MRI experiments.

Supplementary Data 3 –DCE-MRI data sets arising from a longitudinal study of tumor response to anti-angiogenic therapy.